**2020-08-12**

# EXPLORATION STRATEGY FOR THE OUTER PLANETS 2023-2032: Goals and Priorities (UPDATED)

Outer Planets Assessment Group White Paper

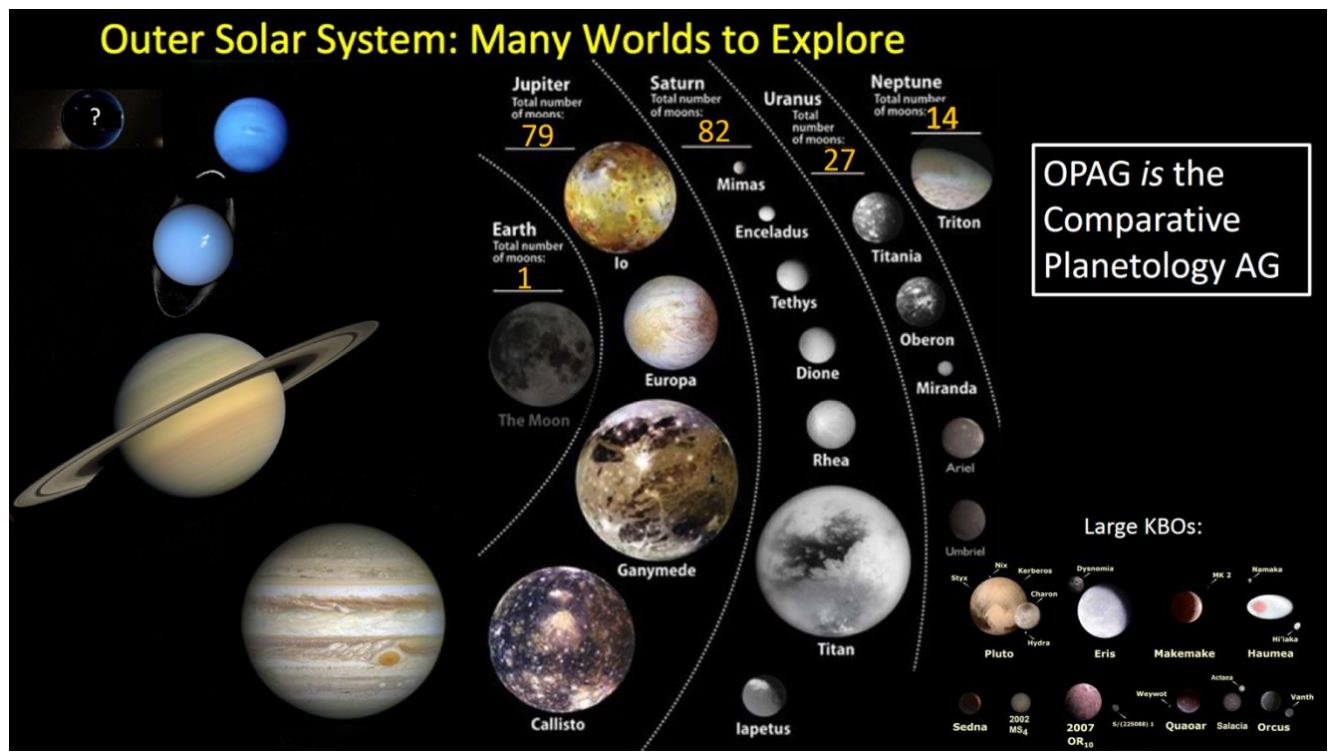

Jeff Moore, Co-Chair, 650-245-5428, jeff.moore@nasa.gov, NASA Ames Research Center
Linda Spilker, Co-Chair, 818-636-0819, linda.j.spilker@jpl.nasa.gov, Jet Propulsion Lab

| | |
|---|---|
| Jeff Bowman | Scripps Institute of Oceanography |
| Morgan Cable | Jet Propulsion Laboratory |
| Scott Edgington | Jet Propulsion Laboratory |
| Amanda Hendrix | Planetary Science Institute |
| Mark Hofstadter | Jet Propulsion Laboratory |
| Terry Hurford | NASA Goddard Space Flight Center |
| Kathleen Mandt | Johns Hopkins APL |
| Alfred McEwen | LPL, University of Arizona |
| Carol Paty | University of Oregon |
| Lynnae Quick | NASA Goddard Space Flight Center |
| Abigail Rymer | Johns Hopkins APL |
| Kunio Sayanagi | Hampton University |
| Britney Schmidt | Georgia Institute of Technology |
| Thomas Spilker | Independent Consultant |

## I. Introduction: The Outer Solar System

The outer solar system is home to a diverse range of objects, holding important clues about the formation and evolution of our solar system, the emergence and current distribution of life, and the physical processes controlling both our own and exoplanetary systems. This White Paper argues for a balanced program to explore this immense region. A combination of directed and competed missions in the coming decade (2023-2032), along with Earth-based efforts, can advance all priority science goals identified by the OPAG community. Those goals are also central to NASA's Strategic Plan.

To organize our priorities, the OPAG community identified three high-level science questions: What is the distribution and history of life in the solar system? What is the origin, evolution, and structure of planetary systems? What present-day processes shape planetary systems and how do these processes create diverse outcomes? Each question is important, and progress should be made on all three. Taking into account the science to be achieved, the timing of solar system events, technological readiness, and programmatic factors, our mission recommendations are as follows. OPAG strongly endorses the completion and launch of the Europa Clipper mission, maintaining the science capabilities identified upon its selection, and a Juno extended mission at Jupiter. For the decade 2023-2032, OPAG endorses a new start for two directed missions: first, a mission to Neptune or Uranus (the ice giants) with atmospheric probe(s), and second, a life detection Ocean World mission (Fig. 1). A Neptune mission is preferred because, while the Neptune and Uranus systems provide equally compelling opportunities to address the Origins and Processes Questions (see Section II), Triton is a higher priority ocean world target in the Roadmaps to Ocean Worlds (ROW) report than the Uranian satellites. The mission to Neptune or Uranus should fly first because a delay threatens key science objectives (see Section III), and additional technological development is required for a directed life detection mission. Thus, while moving forward on Europa Clipper and a mission to Neptune or Uranus, NASA should continue development of life detection technologies, allowing for a new start – before the decade is out – of a dedicated life detection mission in the outer solar system.

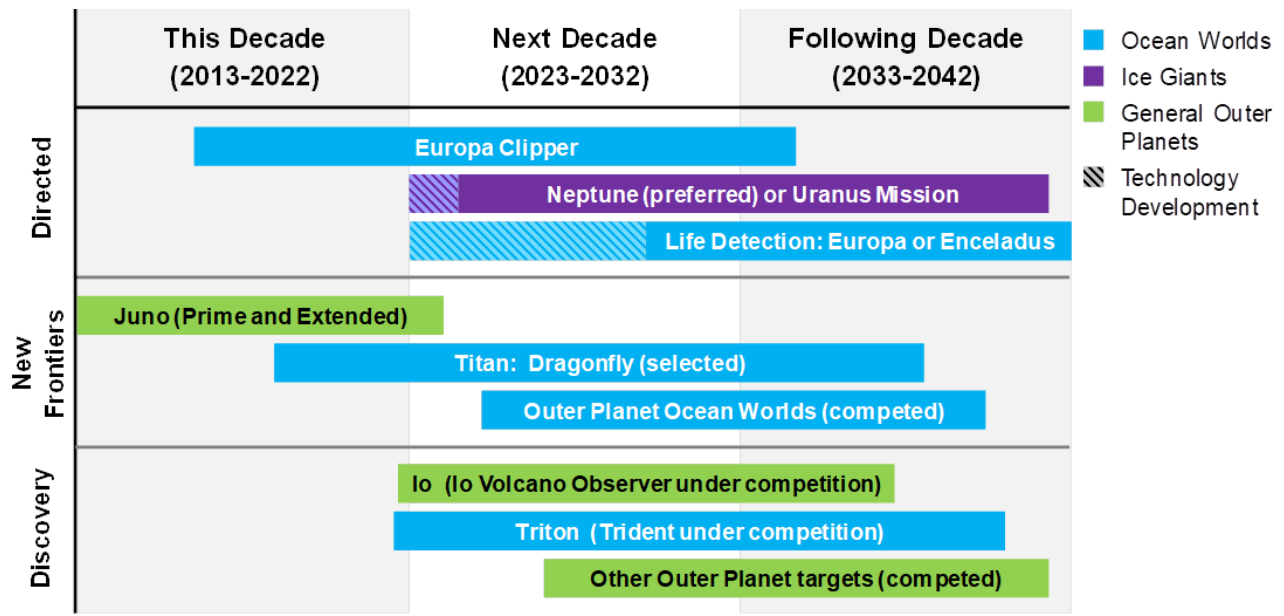

Figure 1: Timeline for a robust Outer Planets Program spanning three decades.

Along with these directed missions, OPAG's science goals extend to all classes of missions but have unique operational challenges, so OPAG encourages making the New Frontiers programs open to all outer planet targets that address Decadal priorities. Along with missions, we emphasize the necessity of maintaining a healthy Research and Analysis (R&A) program which includes a strong laboratory measurements component, as well as a robust Earth-based observing program. International partnerships should continue to be key components of both missions and R&A. OPAG's top two technology priorities are rapid development of a next-generation radioisotope power source for a mission to Neptune or Uranus, and development of key life detection technologies in support of an Ocean World mission.

Finally, fostering an interdisciplinary, diverse, equitable, inclusive, and accessible community is of top importance to the OPAG community. We advocate for continued effort to improve the representation of underrepresented and marginalized people in planetary science.

## II. Fundamental Science Questions and Outer Solar System Mission Concepts

At the request of NASA's Planetary Science Division Director, OPAG undertook a recent survey of its community and devised a set of three "Big Questions" and one unifying/cross-divisional theme for the upcoming Planetary Decadal Survey. Table 1 lists each "Big Question" along with high-level, OPAG-specific sub-questions; the order of Big Questions does not imply priority. These science questions and sub-questions are consistent with the OPAG Goals Document and the Roadmap to Ocean Worlds (ROW) document, both of which were developed with substantial community input. OPAG encourages the Decadal Survey to promote destinations, missions, technologies, and supporting strategic investments that address these questions, particularly as they relate to outer solar system investigations.

**Life Question: What is the distribution and history of life in the solar system?** One of the primary opportunities afforded by outer solar system exploration is the chance to explore subsurface oceans. Oceans may be the key to understanding the origin(s) and evolution of life. The outer solar system is replete with ocean worlds including Europa, Ganymede, Callisto, Enceladus, Titan, and possibly Triton and others. In the inner solar system, only Earth has an ocean today. Ocean worlds are the best places to search for extant life beyond Earth.

OPAG strongly endorses the completion of the Europa Clipper mission. Europa Clipper will provide the requisite knowledge for a future in-situ search for life by investigating the habitability of this ocean world just as the Cassini mission provided required knowledge for future exploration of Titan (i.e. Dragonfly) and Enceladus. Based on Cassini datasets, Enceladus is now considered to be habitable, providing all the necessary ingredients for life as we know it, and has an accessible ocean beneath its ice shell. Technology development in the coming years along with the already planned and perhaps future competed missions to ocean worlds will position NASA to start a directed life detection mission before the end of the decade. We strongly recommend that the next Decadal Survey include a Priority Question about life and/or biosignature detection in addition to habitability studies.

**Origins Question: What is the origin, evolution, and structure of planetary systems?** The prevailing hypothesis of solar system evolution suggests that the outer planets played a pivotal role in molding our solar system through a complex process that included orbital migration of the giant planets, the scattering of planetesimals into the inner and outermost solar system, and the delivery of water and other materials that are essential for life to the terrestrial planets. However, studies of the ice giants (Uranus and Neptune) have revealed limitations in our understanding of basic planetary formation processes. Models predict that ice giants should be rare, yet exoplanet surveys find that they are abundant. Studying the composition and internal structure of all the giant planets provide insights into how, when, and where they formed. Critical measurements (e.g., noble gases and isotopic ratios) require in-situ measurements by an atmospheric entry probe. While the Jupiter and Saturn systems have had dedicated orbiter missions (e.g. Galileo, Juno, Cassini), and an atmospheric probe has been sent into Jupiter, Neptune and Uranus have never had an orbiter mission. Given the importance of dedicated orbiter missions in understanding the origins of our solar system, exploration of ice giants was a top recommendation in the previous Decadal Survey (Vision and Voyages, 2011), and is OPAG's first new directed start recommended for the decade. While both the Neptune and Uranus systems are compelling scientific targets, critical differences exist between them. Ultimately, both must be explored if we are to understand ice giants as a class of planet. However, because Triton has been identified as a higher-priority ocean worlds target than any Uranian satellite, OPAG favors a Neptune mission first.

**Table 1: OPAG Big Questions**

| **Outer Planets Assessment Group (OPAG) Big Questions** |
| --- |
| **Life Question: What is the distribution and history of life in the solar system?** |
| – Does life or do habitable conditions exist beyond the Earth? |
| – What controls the habitability of ocean worlds? |
| – Do ocean worlds host life now, or did they in the past? |
| – What is the potential for prebiotic chemistry in ocean worlds, and how far towards life has this progressed? |
| – What role did the giant planets play in the emergence of life on Earth or elsewhere in the solar system? |
| **Origins Question: What is the origin, evolution, and structure of planetary systems?** |
| – What was the initial chemical profile of the protoplanetary disk as informed by noble gas content in the giant planets, and how did this profile impact the overall formation and evolution of our solar system? |
| – What are the possible architectures of planetary systems, and how do these different configurations affect planet formation and evolution (e.g., giant planet migration, tidal evolution, etc.)? |
| – What controls the formation, evolution and internal structures of gas giants, ice giants, planetary satellites (particularly ocean worlds), rings, and small bodies in the outer solar system? |
| – How do planetary crusts/cryospheres, oceans, atmospheres, and magnetospheres form and evolve in the outer solar system, and how do they influence the evolution of bodies in those systems? |
| **Processes Question: What present-day processes shape planetary systems, and how do these processes create diverse outcomes within and across different worlds?** |
| – How do the chemical and physical processes in the solar system scale between planet size and location within the solar system? |
| – What is the dynamic relationship between the planets, rings, and moons of giant planet systems, and how do these relationships influence their constituent members? |
| – How do the magnetospheres of gas and ice giants influence the dynamics, composition and structure of the atmospheres, rings, and moon surfaces? |
| – How do the aurorae and induced magnetic fields of ocean worlds characterize the coupling between planets, moons, and magnetospheres? |
| – What are the mechanisms, drivers, and rates for transporting heat and materials within, and ejecting them from, (cryo-)volcanically active worlds? |
| – How does coupled orbital evolution and tidal heating affect the interior structures and activity of satellites, and how does the interior evolution of the primaries affect this evolution (e.g., resonance locking)? |
| – What drives the transport of energy and materials within the deep interior of the giant planets? |
| – How do the atmospheric dynamics, cloud microphysics, radiative transfer, and chemistry interact to form stable and transient features observed in outer planet and satellite atmospheres? |
| – How do the ice giant magnetospheres and atmospheres respond to the impulsive solar wind forcing created by their unusual geometries, and what effect does solar insolation play on weather and upper atmospheric structure? |

**Processes Question: What present-day processes shape planetary systems, and how do these processes create diverse outcomes within and across different worlds?** The tremendous diversity of bodies in the outer solar system provides the opportunity for a wide variety of scientific investigations. The satellites of the giant planets, some comparable in size to terrestrial planets, and the dwarf planets of the Kuiper Belt (KBO planets) offer opportunities to study extreme environments on worlds that have experienced very different histories. Tidal heating of satellites leads to current activity and conditions potentially favorable to habitability. The rings and magnetospheres of the giant planets illustrate currently active processes (of collisions and momentum transfer) that played important roles in early stages of solar system formation. The volcanism of Io and the atmosphere of Titan inform important processes on the terrestrial planets and exoplanets. The vast dynamic atmospheres of all four giant planets also serve as natural laboratories to understand fundamental meteorological processes, which are applicable to other planets including Earth.

**Cross-Divisional Theme: How can knowledge of the solar system advance our understanding of the Earth, Sun, and Exoplanets?** This question spans NASA's Planetary, Heliophysics, and Astrophysics Divisions. Specific questions highlighted by the OPAG community are: How does the study of our planet inform our understanding of the outer planets and their moons? How do studies of the diverse present-day oceans in the solar system advance biological, chemical and physical oceanography? How do studies of solar wind interactions at bodies in the outer solar system improve our understanding of the Sun and the propagation and evolution of its dynamic atmosphere? How can solar system bodies inform our understanding of bodies in exoplanetary systems? All missions discussed in this document support the OPAG Cross-Divisional Theme.

**III. Future Missions for Outer Planet Exploration**

The science rationale for OPAG's highest-priority missions was discussed in the previous section. As NASA completes and launches Europa Clipper, OPAG calls for maintaining all science capabilities identified at its selection. As the first new start for a directed mission, OPAG endorses a Neptune or Uranus mission with an atmospheric probe. This mission is of high value to the OPAG community and engages the planetary interiors, atmospheres, rings, small bodies, icy satellite, and magnetospheric disciplines, while also advancing all three priority science questions identified in Section II. Flying to either Neptune or Uranus is scientifically compelling, but Neptune is preferred since Triton is a higher-priority ocean worlds target than the Uranian satellites. We note that **not starting a Neptune or Uranus mission this decade threatens key science objectives**, such as testing the hypothesis that sunlight on Triton's South Polar Cap drives the plume activity seen by Voyager, and – if Uranus is

targeted – exploring a different season and magnetospheric geometry than was seen by Voyager in 1986 and imaging the northern hemispheres of the Uranian satellites. An early start for a Neptune or Uranus mission also maximizes the possible science payload as it enables launch in the optimal 2029—2034 time frame. After a Neptune or Uranus mission, OPAG endorses a directed mission capable of the search for life on an Ocean World, most likely Europa or Enceladus. That mission requires continued technological development of instruments and supporting systems (e.g. sampling) and should be pursued in parallel with the Neptune or Uranus new start. Such technology development could also enable life detection missions to targets outside of the outer planets, such as Mars and Ceres. OPAG strongly recommends that the next Decadal Survey include a Priority Question about life detection, and that Ocean Worlds mission concepts be evaluated in this context given the compelling nature of extant oceans in the outer solar system.

For **New Frontiers** class missions, OPAG supports efforts to find a synergistic solution to maintaining the New Frontiers program's targeted approach to the highest science goals of NASA and the science community, while maintaining the flexibility to respond to new discoveries or new technological opportunities that will evolve in the coming decade. These solutions could be structured in such a way as to allow proposers to respond with creative approaches not envisioned within the survey process, up to and including opening competition to all Ocean Worlds, or to all solar system destinations that address high priority questions advanced in the Decadal Survey. New Frontiers can help address congressional and public interest in an Ocean Worlds program. In all cases, we support the continuing exploration of Ocean Worlds within New Frontiers beyond the selection of the Dragonfly mission to Titan, and advocate for the inclusion of an Enceladus Ocean World mission along with Io Observer (unless selected in the ongoing Discovery competition) and Saturn probes (depending on the outcome of New Frontiers 5). Advances in technology may allow for competed life detection missions or Ice Giants missions within the New Frontiers program. Other concepts deserve consideration as well, such as missions to KBO planets. Given the abundance of worlds to explore in the outer solar system, and the operational challenges of such missions, target restrictions are particularly onerous for the outer planets community.

## IV. Earth-Based Activities

**Research and Analysis (R&A)**

R&A programs are at the core of NASA's scientific research. These programs fund in full or in part the vast majority of all of the U.S.'s outer planet researchers. R&A encompasses telescopic and Earth-based observations, laboratory measurements and experiments, field work, computer simulation and theory, all of which combine to help us understand the origin, evolution, and destiny of planets and satellites. These results are the fundamental products that drive NASA's future planetary missions. A healthy R&A program is essential to ensuring that our missions make the most useful and scientifically valuable measurements, and are thus key to mission success.

**Earth-based Astronomy**

Earth-based observations of outer solar system bodies play an important role in the present and future goals of NASA. These observations, which include those made by space telescopes, sub-orbital missions (including long duration balloon experiments), and ground-based telescopes, have been enormously productive for planetary science at low cost to NASA's Solar System Exploration program, and have provided critically important measurements that have complemented deep space missions. While many of these programs are funded by other divisions at NASA, it is important that the scientific merit of outer solar system observations be emphasized as a key element in the Decadal Survey. It should also be noted that sounding rocket and suborbital research programs provide a unique combination of cost, flexibility, risk tolerance, and support for innovative solutions that make them ideal for the pursuit of unique scientific opportunities, the training of new instrumentalists and PIs, the development of new technology, and infrastructure support.

**Laboratory Measurements and Field Work**

Laboratory measurements and field work provide the "ground truth" for interpreting a wide range of data sets. Without them some observations can be undecipherable, or worse yet, misinterpreted. As our space-borne instruments advance, and as we probe deeper into extreme environments on other planets, there are new needs for such ground truth measurements. For example, many observed spectral features remain unidentified, extrapolated mechanical or rheological properties of materials cannot explain observed geophysical behavior, and sources of observed atmospheric radio opacity are unknown beyond educated guesses. Similarly, the search for Earth-analogs of geologic features on other planets, and testing instruments and techniques in those environments, is crucial.

**International Partnerships**

The science community and NASA have benefited tremendously from international cooperation. Recent missions such as Cassini-Huygens and Juno have taken advantage of substantial foreign contributions, enabling NASA to jointly explore Titan in-situ (via ESA's Huygens Probe delivered by Cassini) and to more cost-effectively explore the Jupiter system, respectively. NASA's relatively modest investments in foreign missions such as ESA's Mars Express and Rosetta have led to remarkable discoveries with contributions by U.S. planetary scientists. Cooperation between space agencies allows the best technical minds across the world to become engaged and this results in better measurements, instruments, and analysis techniques, which add balance to the overall exploration program. Effective international involvement is strongly encouraged in the planning, development, and analysis phases of all space missions to the outer solar system, beginning at the earliest stage possible.

**V. Technology**

Exploration of the outer solar system has challenges that call for technologies at or near the forefront of current capabilities. At this time, there is no radioisotope power system (RPS) suitable for a long-duration outer solar system orbiter, making that a top technological priority. The only RPS currently under active development is NASA's NextGen RTG, and the schedule for that development might not accommodate launch opportunities for high-priority outer solar system missions. The goal of detecting life on an Ocean World (or any other planet in our solar system) is also an open challenge which is under significant development. Issues could arise from investigations that are insufficiently agnostic (i.e. too 'Earth-centric'), and measurements may be complicated by interfering species or sampling limitations. Optimal biosignature detection strategies will likely involve multiple, independent tests for life, such that all possible abiotic explanations for the results may be effectively ruled out. OPAG strongly encourages continued investment in development of life detection instruments, as well as supporting technologies (sampling systems, contamination control, planetary protection, etc.) to address this 'civilization-level' science in the next decade.

The breadth of technology needed for outer planet exploration calls for an aggressive and focused technology development strategy that aligns with the Decadal Survey's recommended mission profile, and that includes technologies developed by NASA as well as acquisition of applicable technologies from other government and commercial sectors. OPAG specifically advocates for a focused technology program for a directed mission to Neptune or Uranus to be ready for a launch in the late-2020s. A full discussion of technology needs can be found in a companion white paper (T. Spilker, et al.).

**To explore the outer solar system requires advanced technology. OPAG recommends the following to enable this exploration:**
- NASA should work with the relevant agencies to complete the development of the full 1.5 kg/year Pu-238 production capacity on a schedule consistent with providing needed material for future outer planet missions.
- A focused technology program for a directed mission to Neptune or Uranus should be initiated in order to be ready for a launch in the late 2020s. This must include a next-generation radioisotope power system suitable for a long-duration orbiter.
- NASA should expand the funding of advanced communication and radio science technologies required for outer planets, especially increasing data rate capabilities from Neptune or Uranus.
- NASA should continue to invest in the development of underlying technologies (thrusters, power and control, propulsion technologies) for solar-electric propulsion.
- NASA should invest in aerocapture technologies and implement incentives for its use at less-challenging destinations such as Venus or Titan, paving the way for future use at the Giant Planets.
- For planetary probes, OPAG recommends investment in maintaining heritage and developing alternative thermal protection system (TPS) materials.
- NASA should invest in life detection technologies suitable for use on an Ocean World, such as instruments, sampling systems, contamination control and planetary protection.

**VI. Fostering an Interdisciplinary, Diverse, Equitable, Inclusive, and Accessible Community**

We advocate for the need to continue efforts to ensure equitable and inclusive representation in the planetary exploration endeavor. Studying the outer solar system requires a highly interdisciplinary and collaborative approaches because it contains every category of planetary bodies with a wide range of physical, chemical and potentially biological processes. We advocate for crafting future mission and research opportunities to improve representation of groups who are underrepresented and marginalized in planetary science and society as a whole.

**VII. Summary**

Exploring the outer solar system is difficult, but immensely rewarding. Voyager, Galileo, Cassini-Huygens, and New Horizons rank among humankind's great voyages of exploration. The OPAG Goals Document outlines an exciting course to continue this epic exploration in the upcoming decade. For large directed missions, OPAG endorses two new missions: first, a mission to Neptune or Uranus, followed by a life detection Ocean World mission. A next-generation RPS system and advances in life detection technology are needed for those missions, respectively, to proceed. For New Frontiers class missions, OPAG supports opening competition to all missions addressing the top goals of the Decadal Survey. If targets are restricted, an open call for Ocean Worlds missions should be included alongside the NF5 list of targets. Finally. we recommend the decadal survey to encourage the planetary science community to foster an interdisciplinary, diverse, equitable, inclusive, and accessible environment.

**Key References**

NASA 2017 Ice Giant Study Report: https://www.lpi.usra.edu/icegiants/mission_study/

Roadmaps to Ocean Worlds (ROW) study report: https://www.lpi.usra.edu/opag/ROW/

OPAG Science Goals document, https://www.lpi.usra.edu/opag/goals-08-28-19.pdf